%% file: main.tex
\documentclass{sig-alternate-05-2015}
\usepackage{color}

\begin{document}

% Copyright
%\setcopyright{acmcopyright}
\setcopyright{rightsretained}
%\setcopyright{acmlicensed}
%\setcopyright{rightsretained}
%\setcopyright{usgov}
%\setcopyright{usgovmixed}
%\setcopyright{cagov}
%\setcopyright{cagovmixed}

% DOI ???
%\doi{}

% ISBN ???
%\isbn{}

%Conference
\conferenceinfo{Neu-IR '16 SIGIR Workshop on Neural Information Retrieval,}{July 21, 2016, Pisa, Italy}

%\acmPrice{\$15.00}

%
% --- Author Metadata here ---
%\conferenceinfo{Neu-IR '16 SIGIR Workshop on Neural Information Retrieval}{July 21, 2016, Pisa, Italy}
%\CopyrightYear{2007} % Allows default copyright year (20XX) to be over-ridden - IF NEED BE.
%\crdata{0-12345-67-8/90/01}  % Allows default copyright data (0-89791-88-6/97/05) to be over-ridden - IF NEED BE.
% --- End of Author Metadata ---

\title{Adaptability of Neural Networks on Varying Granularity IR Tasks}
%\subtitle{[Extended Abstract]}
%\titlenote{A full version of this paper is available as
%\textit{Author's Guide to Preparing ACM SIG Proceedings Using
%\LaTeX$2_\epsilon$\ and BibTeX} at

\numberofauthors{1} %  in this sample file, there are a *total*
\author{
% 1st. author
% UNCOMMENT FINAL DRAFT
\alignauthor
Daniel Cohen, Qingyao Ai,  W. Bruce Croft\\
      \affaddr{Center for Intelligent Information Retrieval}\\
      \affaddr{College of Information and Computer Sciences}\\
      \affaddr{University of Massachusetts Amherst}\\
      \affaddr{Amherst, MA}\\
      %\affaddr{College of Information and Computer Sciences, \\ University of Massachusetts Amherst, Amherst, MA, USA}
      \email{\{dcohen, aiqy, croft\}@cs.umass.edu}\\
}
% There's nothing stopping you putting the seventh, eighth, etc.
% author on the opening page (as the 'third row') but we ask,
% for aesthetic reasons that you place these 'additional authors'
% in the \additional authors block, viz.
%\date{11 February 2016}
% Just remember to make sure that the TOTAL number of authors
% is the number that will appear on the first page PLUS the
% number that will appear in the \additionalauthors section.

\maketitle
\begin{abstract} 
Recent work in Information Retrieval (IR) using Deep Learning models has yielded state of the art results on a variety of IR tasks. Deep neural networks (DNN) are capable of learning ideal representations of data during the training process, removing the need for independently extracting features. However, the structures of these DNNs are often tailored to perform on specific datasets. In addition, IR tasks deal with text at varying levels of granularity from single factoids to documents containing thousands of words. In this paper, we examine the role of the granularity on the performance of common state of the art DNN structures in IR.
\end{abstract}
%comments for above
% non-factoid should probably be hyphenated (non-factoid)
% I feel that there are too many 'relations' in sentence starting with Neural networks are capable..., perhaps use a different word
% short-term should be hyphenated





%
%  Use this command to print the description
%

% We no longer use \terms command


\input{Introduction}

\input{Related_work}
\input{Granularity_Tasks}
~\\
\input{Conclusion}

%\section{Related Work}
\section{Acknowledgment}
This work was supported in part by the Center for Intelligent Information Retrieval and in part by NSF IIS-1160894. Any opinions, findings and conclusions or recommendations expressed in this material are those of the authors and do not necessarily reflect those of the sponsor.

%%%%%%%%%%%%%%%%%%%%%%%%%%%%%%%%%%%%%%%%%%%%%%%%%%%%%%%%%%%%%%%%%%%%%%

%\end{document}  % This is where a 'short' article might terminate

%ACKNOWLEDGMENTS are optional

\bibliographystyle{abbrv}
\bibliography{sigproc}  % sigproc.bib is the name of the Bibliography in this case
% ACM needs 'a single self-contained file'!
%\balancecolumns
%\balancecolumns % GM June 2007
% That's all folks!
\end{document}

%% file: Introduction.tex
\section{Introduction}

Learning effective representations of data is a critical component of any system that ranks documents. Conventional approaches rely on transforming text into vectors consisting of lexical, semantic, and syntactic features that capture the information contained in text. This conversion depends on domain knowledge and is an independent step from the optimization process of the ranking method. As this process is separate from the loss function, potential information can be lost that negatively affects performance.
Deep learning has been shown to learn internal representations directly from the text in natural language processing and specific IR tasks that yield state of the art performance. However, the deep learning models used for these IR tasks are often tailored for the individual task with the network structure making some assumption about the data, and little work has been done in examining how well these networks can adapt to collections with varying levels of granularity.

IR focuses on retrieval of information at differing levels of granularity whether at the single word level in the factoid task such as TREC~QA, passage level involving community question and answer, or the document level dealing with ad-hoc retrieval. Each of these levels present unique challenges when fitting a model, and we show that DNNs are not exempt from this problem. In the following sections we examine state of the art DNNs on varying levels of granularity to demonstrate the efficacy of different neural structures at each level of granularity.

%% file: Related_work.tex
\section{Related Work}
At each level of granularity, significant improvements have been made by introducing various DNN structures. Convolutional neural networks (CNN) have been used at various layers in the neural net, at the input level over word embeddings demonstrated by Severyn and Moschitti~\cite{sev2015cnn}, as an intermediary layer within a feedforward network introduced by Feng et al.~\cite{feng2015}, or as a penultimate stage on top of a recurrent neural network (RNN) to provide more composite representations over the question and answer text by Tan et al.~\cite{tan2015lstmcnn}. Regardless of the position of the convolutional layer, the motivation behind implementing a convolutional layer was to extract the most salient features from the input to allow easier similarity comparisons.

As language is sequential in nature, RNNs have been shown to work extremely well for IR tasks. Wang and Nyberg~\cite{wang2015blstm1,wang2015blstm2} show that using a bidirectional Long Short-Term Memory (BiLSTM) network over query-answer pairs to determine relevance is an effective approach to the fine grain level of the TREC~QA task as well as for passage level retrieval. Providing additional insight to how LSTM networks process text, Palangi et al.~\cite{palangi2016lstm} demonstrate the use of a weakly supervised LSTM network to detemine answer sentence similarity and examine how individual cells attenuate information when processing query-answer pairs.

\iffalse
Severyn and Moschitti~\cite{sev2015cnn} have shown the effectiveness of convolutional neural networks on the short granularity task of the TREC~QA and TREC microblog tracks. The network they implemented involved a convolutional layer to extract salient features. A subsequent matrix then computes the similarity between the two representations and is concatenated with individually pooled representations of the sentences. This representation is then combined with term frequency statistics prior to a dense layer. This network achieved state of the art results on the TREC~QA track, outperforming other DNN networks.
Wang et al.~\cite{wang2015blstm} investigate the use of bidirectional long short term memory (BiLSTM) networks on the same TREC~QA task. They utilized two bidirectional LSTM layers which sequentially read in the query concatenated with candidate sentences. The LSTM network does not outperform the CNN implementation, although it does come within 1\% and 3\% of the previous CNN network~\cite{sev2015cnn}. 
\fi

Another important attempt that applies DNN for IR tasks is the Deep Structured Semantic Model (DSSM) and its variations. 
Introduced by Huang et al.~\cite{huang2013learning}, DSSM uses a word hashing technique to project varied length text into fixed length vectors as the model's input and constructs a feed-forward neural network above it. The relevance between documents and queries is measured by the cosine similarity between their output vectors. 
Recently, Shen et al.~\cite{shen2014latent} proposed a convolution network (CLSM) and Palangi et al.~\cite{palangi2014semantic} proposed a RNN-LSTM model with the same word hashing technique of DSSM. They showed positive results when applying these models on ad-hoc retrieval tasks with web page titles collections. 
However, the effectiveness of word hashing and DSSM models vary considerably as text length changes. 
Their performance on standard TREC collections are still poor according to our experiments.

%% file: Granularity_Tasks.tex
\section{Granularity Tasks}
We examine the efficacy of deep learning on three distinct levels of granularity. First, at the fine grain level, retrieval focuses on a specific  word or deals with a short sentence of text containing the relevant information. Second, at the medium granularity level, the information need of the query can no longer be addressed by a single sentence, and often requires multiple sentences to be relevant. Third, we address the coarse grain level, which we view as full document retrieval commonly found on ad-hoc retrieval tasks.
\input{Short_task}

\input{Passage_task}

\input{Long_task}

%% file: Short_task.tex
\subsection{Fine Granularity}
The focus of this section is on the TREC~QA task. In this task, the length of individual documents are often no more than a single sentence, and queries consist of short questions such as \textit{``When did James Dean die?"} or \textit{``What is crips' gang color"}. The relevant information in each document is one or two words that directly address the information need of the query.

From the deep learning perspective, CNNs adapt well to the fine grained task as they are able to identify key aspects of an input matrix. This ability has resulted in these networks receiving widespread use in the computer vision task. The same principle can be applied to the sentence level by allowing convolutional layers to extract the most salient information over embeddings of a sentence.
This approach has been used for semantic sentence level matching by Hu et al.~\cite{hu2015cnn}. Severyn and Moschitti~\cite{sev2015cnn} also take advantage of the matching ability of CNNs by implementing a convolutional layer to extract the most salient features between answer and query sentences to compute similarity scores for ranking.

An interesting note is the performance of RNNs at the same granularity level. As shown in Yin et al.~\cite{yin2016abcnn} and Santos et al.~\cite{santos}, conventional CNNs often outperform equivalent LSTM networks at this level of granularity as filter lengths are able to capture the language dependencies and match keywords when the candidate answer sentences are short.

%that due to the length of this task, RNNs fail to outperform CNNs when reranking sentence length documents. This can be attributed to the nature of the TREC~QA collection, where answers are specific words without any long term dependencies across the sentence. Thus RNNs have little additional information to gain by processing the sentence sequentially when compared to CNNs, as seen by Wang and Nyberg's Bidirectional LSTM model~\cite{wang2015} in Table~\ref{tab:trecqa}.

%% file: Passage_task.tex
\subsection{Medium Granularity}
\begin{table}
    \centering
    \begin{tabular}{c || c | c }
        Method & MRR & P@1 \\\hline\hline
		LSTM & 0.6314 & 0.7849 \\ \hline
		%DSSM & 0.095 & 0.201 & 0.171 \\ \hline 
		%CLSM & 0.067 & 0.146 & 0.125 \\ \hline
        CNN & 0.3729 & 0.6225 \\ \hline
    \end{tabular}
    \caption{Comparison of a CNN and LSTM network after hyper-parameter tuning over Yahoo's Webscope L4 collection.}
    \label{tab:L4}
\end{table}
The medium granularity level, consisting of passages, contrasts sharply with the granularity of the previous section. Instead of identifying specific words contained within a sentence, the passage task deals with information related to the query that can span multiple sentences. However, relevance is not determined solely by topical similarity between document and query. Text in relevant passages can have little term overlap with the query, and conventional IR methods such as BM25 have reflected this in their performance.

Due to the span of relevant information across the length of candidate answers,
LSTM networks are uniquely suited to this task as they are able to model syntactic and semantic dependencies across positions in a sequence and focus less on matching than the CNN does. We demonstrate this on Yahoo's Webscope L4 CQA collection~\cite{yahoo} of ``manner'' type questions, where a LSTM model built in a similar fashion to ~\cite{wang2015blstm1} significantly outperforms an equivalent CNN network as shown in Table~\ref{tab:L4}. The purpose of this test was to demonstrate the ability of the two network structures to retain information across long sequences, therefore the candidate answer pool for each query consisted of 10 randomly sampled answers from the collection. These candidate answers are significantly longer than those found in WikiQA~\cite{feng2015} or the TREC~QA task with a mean length of 75. The filter lengths of CNNs are unable to capture long term dependencies that span multiple sentences, which results in its poor performance relative to the LSTM network.

Palangi et al.~\cite{palangi2016lstm} investigate the internal representation of text within a LSTM network. Internal cell states accumulate semantic information across sequences, and their corresponding inputs learn to respond to semantically related words specific to each cell. In addition, the LSTM network is capable of keyword recognition to directly match query and document similarities. This contrasts with a standard RNN without LSTM cells, where the length of the passage task results in the internal representation `forgetting' previous information due to the vanishing gradient problem.

%% file: Long_task.tex
\subsection{Coarse Granularity}

%what specific to ad-hoc retrieval
%what is it effect on deep learning model
% varied length -> training problem

% complicated topic distribution -> fixed size vector representations would lose finer granularity
Tasks with coarse granularity including Ad-hoc retrieval are usually concerned with collections of text with great variation in length. 
Although the queries tend to be shorter, the documents range from tens of words to thousands of terms.
Accompanying the challenge that length variation poses, the concept of relevance varies from document to document as the relevant portion of a document might range from a few sentences to its entirety. These two unique properties of coarse granularity collections result in different challenges for DNNs which are not apparent at other granularity levels.

%Given two relevant documents, the entirety of the first document can be related to the query and the second document might only have a few sentences that are relevant.

%are different concepts of relevance and different challenges for deep neural networks distinct from fine and medium granularity levels discussed in the previous sections.
We applied conventional networks discussed in Section 3.2 which performed well on passage length answers, but were unable to perform better than random over the Robust04 collection and thus require a different approach.  

\textbf{Varied input length.}
%DSSM
In ad-hoc retrieval tasks, the length differences in documents are so large that they significantly affect the training of deep models. 
Most neural models include a step that converts varied length input into fixed length vectors (i.e. input layer for DNN, pooling in CNN and memory vectors in RNN).
Without accounting for the original length of text, this process could introduce strong bias for short or long documents.
For example, to train a deep structure semantic model (DSSM) for ad-hoc retrieval, Huang et al.~\cite{huang2013learning} proposed a word hashing technique that aggregates n-grams of terms to produce a fixed length representation for each document.
When applied to short text like web page titles, which have few n-grams, word hashing produces high quality representations without losing too much information.
However, this technique becomes problematic as document length increases from tens of words to hundreds of words.
According to our observations, the n-gram representations for documents with hundreds of words are dense and noisy. 
%Consequentially, DSSM shows different discriminative ability in documents with different length.
%Huang et al.~\cite{huang2013learning} reported impressive performance of DSSM in experiments with web page titles, but 
Unsurprisingly, our experiments with DSSM on standard ad-hoc retrieval collections were not effective.
The MAP of DSSM and its convolution version (CLSM) are less than 0.1 on Robust04 title queries. 
Notice that the same metric for query likelihood (QL) model is 0.253 as shown in Table~\ref{tab:ad_hoc}.

\textbf{Varied relevance granularity.}
\begin{table}
    \centering
    \begin{tabular}{c || c | c | c }
        Method & MAP & nDCG@20 & P@20 \\\hline\hline
		QL & 0.253 & 0.415 & 0.369 \\ \hline
		%DSSM & 0.095 & 0.201 & 0.171 \\ \hline 
		%CLSM & 0.067 & 0.146 & 0.125 \\ \hline
        WE & 0.135 & 0.257 & 0.227 \\ \hline
        PV & 0.177 & 0.288 & 0.264 \\ \hline
        WE-LM & 0.255* & 0.417* & 0.370* \\ \hline
        PV-LM & 0.259* & 0.418 & 0.371 \\ \hline
    \end{tabular}
    \caption{Comparison of different models over the Robust04 collection with title queries. * means significant difference over QL respectively at 0.005 significance level measured by Fisher randomization test.}
    \label{tab:ad_hoc}
\end{table}
%Word embedding, PV
Another problem that makes ad-hoc retrieval difficult for existing deep models is the vague definition for relevance. 
A short document could be relevant to a query because its main topic is related to the query.
Meanwhile, a long document could be relevant to a query if it has a subtopic that describes the query.
This characteristic of ad-hoc retrieval presents challenges to both supervised and unsupervised neural models.
For supervised models like DNN and RNN, the back propagation of relevance information affects the gradient computation on all input words. However, most of these words may not be related to the document's relevance with a specific query (especially for long documents).
A considerable amount of labeled data is needed in order to learn the weights for a model that can understand the relevance of a document from different angles.

For unsupervised models like WE~\cite{vulic2015monolingual} and the paragraph vector model (PV)~\cite{le2014distributed}, the embedding representations of documents are constructed to capture their main topics. 
These representations lack discriminative ability at query time because we cannot distinguish the finer difference between semanticly related words and subtopics~\cite{zhai2008statistical}.
%add experiments
For example, Table~\ref{tab:ad_hoc} shows the performance of retrieval models including WE and PV. 
WE~\cite{vulic2015monolingual} aggregates embeddings of words to form document representations and ranks documents according to their cosine similarities with queries. 
PV estimates a language model with paragraph vector model~\cite{le2014distributed} and ranks documents according to the likelihood of queries given document models.
Using WE and PV solely did not perform well compared to QL with dirichlet smoothing. 
We only achieve positive results when we combined these models with language modeling approaches that explicitly capture the exact matching information of queries and documents. 
As these problems pose significant challenges from a deep learning perspective, one direction of future research is examining the role of the attention mechanism when dealing with documents at the coarse granularity level.

%% file: Conclusion.tex
\section{Conclusion}
We have shown the efficacy and shortcomings of common neural architectures at varying levels of IR task granularity. When candidate answers are short, CNNs and LSTM networks perform at equivalent levels with differences attributed to attention methods and structure differences beyond the convolutional and LSTM layers. At the passage level, we demonstrate that LSTMs are able to store additional temporal information which an equivalent CNN is unable to accomplish. Lastly, we discuss the unique problem that ad-hoc retrieval poses for neural networks, and potential solutions to overcome these issues.